\documentclass[12pt]{iopart}
\usepackage{graphicx}

\begin{document}
\def \ee {\varepsilon}
\thispagestyle{empty}
\title[]{
Generalized plasma-like permittivity and
thermal Casimir force between real metals
}

\author{
B.~Geyer,${}^{1}$
G.~L.~Klimchitskaya${}^{1,2}$
and V.~M.~Mostepanenko${}^{1,3}$
}

\address{${}^1$Center of Theoretical Studies and Institute for Theoretical 
Physics, Leipzig University,
D-04009, Leipzig, Germany}

\address{$^2$North-West Technical University, Millionnaya St. 5,
St.Petersburg, 191065, Russia
}

\address{$^3$
Noncommercial Partnership  ``Scientific Instruments'', 
Tverskaya St. 11, Moscow, 103905, Russia}

\begin{abstract}
The physical reasons why the Drude dielectric function is
not compatible with the Lifshitz formula, as opposed to the
generalized plasma-like permittivity, are presented.
Essentially, the problem is connected with the finite
size of metal plates. It is shown that the Lifshitz theory
combined with the generalized plasma-like permittivity
is thermodynamically consistent.
\end{abstract}

\section{Introduction}

In the last few years the Casimir effect \cite{1} received
common recognition as one of the most important subjects
of interdisciplinary interest. The Casimir force is 
of the same nature as
other one-loop vacuum effects of quantum
electrodynamics \cite{2}. It arises due to the alteration
of the spectrum of electromagnetic zero-point oscillations
by material boundaries. Early
stages of modern experiments and related theory are
reflected in \cite{5}.
Recent trends go towards complex
experimental and theoretical studies of the Casimir effect,
including the applications to nanotechnology.
For this purpose many classical theoretical results on
the subject obtained within the framework of quantum
field theory (see, e.g., monographs \cite{2a,3,4}, 
reviews \cite{5,4aA,4bA}
proceedings \cite{4a} and more recent papers \cite{4b,4bb}) 
should be adapted to the case of real material bodies.
Realistic material properties are important also for applications
of the Casimir effect in nanotechnology \cite{7a,4c,4d}.

The basic theory of the van der Waals and Casimir forces
between dielectric materials was proposed by Lifshitz
\cite{6}. However, the application of this theory to
Drude metals and semiconductors with sufficiently low charge
carrier density met serious problems.
Namely, at first,
it was shown \cite{7,8} that for Drude metals with
perfect crystal lattices the Lifshitz theory violates
the Nernst heat theorem. This is connected with the fact that
the reflection  coefficient for the transverse electric
mode of the electromagnetic field at zero frequency is
equal to zero if the dielectric permittivity behaves as
$\omega^{-1}$ when the frequency $\omega$ vanishes.
For metals with impurities the Nernst heat theorem
is formally preserved \cite{9}. This inclined the proponents
of the Drude model to believe that it is applicable
together with the Lifshitz theory (different
arguments on this problem can be found in \cite{10,11,12,13}).
However, precision measurements of the Casimir force
\cite{14,15,16,17} excluded the Drude model at a 99.9\%
confidence level.

Another problem arises when the Lifshitz theory is applied  to
dielectrics or semiconductors with not too high density
of free charge carriers. In this case the Nernst heat
theorem is violated if the conductivity at zero frequency
is taken into account \cite{18,19,20,21}. Here, it
 is the discontinuity of the transverse
magnetic mode at zero frequency, which is responsible for 
that violation. Furthermore, it was demonstrated 
\cite{22,23} that the inclusion of the conductivity at
zero frequency into the Lifshitz theory leads to a
contradiction with experiment. Until recently there exists
not any theoretical approach to the thermal Casimir force
which would be in agreement with both short separation
experiments \cite{24,25} and longer separation experiments
\cite{14,15,16,17}. The approach using the usual,
nondissipative, plasma model was shown to be in agreement
with longer separation experiment \cite{14,15,16,17}, but
to be in contradiction with the experiment performed at short
separations \cite{24,25}. The impedance approach \cite{26} 
was also found in agreement with longer separation
experiments \cite{14,15,16,17},
but it is simply not applicable at shorter separations characteristic
for the experiment of \cite{24,25}. Because of this, the
Lifshitz theory at zero temperature by necessity was used
for the comparison between the short separation experiment
\cite{24,25} and theory, even though that experiment was 
performed at a room temperature of 300\,K.

Recently, a new theoretical approach to the thermal Casimir
force between real metals has been proposed \cite{27} using
the generalized plasma-like dielectric permittivity.
The latter includes dissipation processes due to the interband
transitions of core electrons but disregards dissipation due to
scattering processes of free electrons. As was shown in
\cite{27}, the Lifshitz formula combined with the generalized
plasma-like dielectric permittivity is consistent with both
short and long separation experiments. It also exactly
satisfies the Kramers -Kronig relations. However, the question
why one should include one type of dissipation (interband
transitions of core electrons) to fit theory to experiment
while disregarding another one (scattering processes
of free electrons) remained unresolved.

In this paper we present and discuss the physical explanation
why the Drude dielectric function cannot be used to describe
the thermal Casimir force between metal plates of 
{\it finite area}. The idea of that explanation was briefly
published first by Parsegian \cite{28}, but did not attract
the attention it is deserving. As we show below, the Drude
dielectric function is not compatible with the
zero-frequency term of the Lifshitz formula if the area
of plates is finite. We also perform a rigorous
analytical proof of the fact that the Casimir entropy
calculated using the generalized plasma-like dielectric
permittivity satisfies the Nernst heat theorem.

The paper is organized as follows. In Section~2 we explain
why the Drude dielectric function is incompatible
with the Lifshitz formula in the case of two parallel
metallic plates of finite area. Section\ 3 contains an
asymptotic derivation of the analytic expression for
the Casimir entropy in the limit of low temperatures.
Here we present a proof for the validity of the Nernst
heat theorem in the Lifshitz theory combined with the
generalized plasma-like dielectric permittivity.
Section\ 4 contains our conclusions and a discussion.

\section{Why the Drude dielectric function is not
compatible with the Lifshitz formula for metallic
plates of finite area}

In the framework of the Lifshitz theory the free energy of the
fluctuating electromagnetic
field between two electrically neutral plane
parallel plates of thickness $d$ at temperature $T$ in
thermal equilibrium is given by \cite{5,6}
\begin{eqnarray}
&&
{\cal F}(a,T)=\frac{k_BT}{2\pi}\sum\limits_{l=0}^{\infty}
\left(1-\frac{1}{2}\delta_{0l}\right)
\int_{0}^{\infty}k_{\bot}\,dk_{\bot}
\nonumber \\
&&\phantom{aaa}
\times\left\{\ln\left[1-r_{\rm TM}^2(\xi_l,k_{\bot})
e^{-2aq_l}\right]+
\ln\left[1-r_{\rm TE}^2(\xi_l,k_{\bot})
e^{-2aq_l}\right]\right\}.
\label{eq1}
\end{eqnarray}
\noindent
Here $a$ is the separation distance between the plates,
$k_B$ is the Boltzmann constant, $\xi_l=2\pi k_B Tl/\hbar$
are the Matsubara frequencies defined for any
$l=0,\,1,\,2,\,\ldots\,$, and 
$k_{\bot}=|\mbox{\boldmath$k$}_{\bot}|$ is the magnitude
of the wave vector projection onto the plane of the plates.
The reflection coefficients for the two independent
polarizations of the electromagnetic field (transverse
magnetic, TM, and transverse electric, TE) are expressed
\cite{29} in terms of the frequency-dependent dielectric
permittivity, $\varepsilon(\omega)$, along the imaginary
frequency axis:
\begin{eqnarray}
&&
r_{\rm TM}(\xi_l,k_{\bot})=
\frac{\varepsilon_l^2q_l^2-k_l^2}{\varepsilon_l^2q_l^2+
k_l^2+2q_lk_l\varepsilon_l\coth(k_ld)},
\nonumber \\
&&
r_{\rm TE}(\xi_l,k_{\bot})=
\frac{k_l^2-q_l^2}{q_l^2+k_l^2+2q_lk_l\coth(k_ld)},
\label{eq2}
\end{eqnarray}
\noindent
where
\begin{equation}
q_l=\sqrt{k_{\bot}^2+\frac{\xi_l^2}{c^2}}, \quad
k_l=\sqrt{k_{\bot}^2+\varepsilon_l\frac{\xi_l^2}{c^2}}, 
\quad \varepsilon_l=\varepsilon({\rm i}\xi_l).
\label{eq3}
\end{equation}

Equation (\ref{eq1}) was originally derived \cite{6} for 
dielectric plates of infinite area. However, it is commonly
used for plates of finite area $S$ under the condition
$a\ll\sqrt{S}$. If this condition is satisfied, corrections
to Eq.~(\ref{eq1}) due to the finiteness of plate area are
shown to be negligibly small \cite{5,30} for both dielectric
and ideal metal plates. Below we show that this is not the
case for  metal plates described by the Drude dielectric
function where the presence of a real current of conduction
electrons leads to a crucially new physical situation.

Papers \cite{9,12,13,31} describe metallic plates by using
the dielectric permittivity of the Drude model,
\begin{equation}
\varepsilon_D(\omega)=1-\frac{\omega_p^2}{\omega(\omega+{\rm i}\gamma)},
\label{eq4}
\end{equation}
\noindent
where $\omega_p$ is the plasma frequency and $\gamma$ is the
relaxation parameter. As is correctly stated by Parsegian
(see \cite{28}, p.254), ``this is valid only in the case of an
effectively infinite medium where no walls limit the flow
of charges.'' To gain a better understanding of this statement,
we derive Eq.~(\ref{eq4}) starting from Maxwell equations in
an unbounded nonmagnetic metallic medium
\begin{eqnarray}
&&
\mbox{rot{\boldmath$B$}}=\frac{1}{c}
\frac{\partial\mbox{\boldmath$E$}}{\partial t}+
\frac{4\pi}{c}\sigma_0\mbox{\boldmath$E$},
\qquad \mbox{div{\boldmath$B$}}=0,
\nonumber \\
&&
\mbox{rot{\boldmath$E$}}=-\frac{1}{c}
\frac{\partial\mbox{\boldmath$B$}}{\partial t},
\qquad \mbox{div{\boldmath$E$}}=0.
\label{eq5}
\end{eqnarray}
\noindent
Here, the electric current density
$\mbox{\boldmath$j$}=\sigma_0\mbox{\boldmath$E$}$ is
induced in a metal under the influence of external sources,
and $\sigma_0$ is the conductivity at zero frequency.
Physically the demand that the medium is unbounded means
that it should be much larger than the extension of the
wave fronts of electromagnetic waves coming from external
sources (i.e., of zero-point oscillations and thermal
photons).

Solutions of equations (\ref{eq5}) can be found in the
form of monochromatic waves,
\begin{equation}
\mbox{\boldmath$E$}=\mbox{Re}\left[\mbox{\boldmath$E$}_0
(\mbox{\boldmath$r$})e^{-{\rm i}\omega t}\right],
\quad
\mbox{\boldmath$B$}=\mbox{Re}\left[\mbox{\boldmath$B$}_0
(\mbox{\boldmath$r$})e^{-{\rm i}\omega t}\right],
\label{eq6}
\end{equation}
\noindent
where $\mbox{\boldmath$E$}_0(\mbox{\boldmath$r$})$ and
$\mbox{\boldmath$B$}_0(\mbox{\boldmath$r$})$ satisfy
equations
\begin{equation}
\Delta\mbox{\boldmath$E$}_0(\mbox{\boldmath$r$})+
k^2\mbox{\boldmath$E$}_0(\mbox{\boldmath$r$})=0,
\quad
\Delta\mbox{\boldmath$B$}_0(\mbox{\boldmath$r$})+
k^2\mbox{\boldmath$B$}_0(\mbox{\boldmath$r$})=0,
\label{eq7}
\end{equation}
\noindent
following from (\ref{eq5}) with
\begin{equation}
k^2=\frac{\omega^2}{c^2}+{\rm i}\frac{4\pi\sigma_0\omega}{c^2}
\equiv\frac{\varepsilon_n(\omega)\omega^2}{c^2}.
\label{eq8}
\end{equation}

Here the dielectric permittivity of the normal skin effect,
$\varepsilon_n(\omega)$, is introduced 
\begin{equation}
\varepsilon_n(\omega)=1+{\rm i}\frac{4\pi\sigma_0}{\omega}.
\label{eq9}
\end{equation}
\noindent
This equation is applicable at not too high frequencies
(the region of the normal skin effect) where the
relation 
$\mbox{\boldmath$j$}=\sigma_0\mbox{\boldmath$E$}$
is valid. The Drude model extends the applicability of
(\ref{eq9}) to higher frequencies, up to the plasma
frequency, by making the following replacement in
equation (\ref{eq9}):
\begin{equation}
\sigma_0{\>}\to{\>}\sigma(\omega)=
\frac{\sigma_0\left(1+\frac{{\rm i}\omega}{\gamma}\right)}{1+
\frac{\omega^2}{\gamma^2}}.
\label{eq10}
\end{equation}
\noindent
Substituting (\ref{eq10}) in (\ref{eq9}) and taking into
account that $\sigma_0=\omega_p^2/(4\pi\gamma)$ \cite{32}
we recover the dielectric permittivity of the Drude model
(\ref{eq4}). At sufficiently high frequencies
$\gamma\ll\omega<\omega_p$ (the region of infrared optics)
one can neglect unity as compared to $\omega/\gamma$
and $\omega^2/\gamma^2$ in (\ref{eq10}). 
Then (\ref{eq4}) and (\ref{eq10})
lead to the so-called free electron plasma model
\begin{equation}
\varepsilon_p=1-\frac{\omega_p^2}{\omega^2},
\qquad
\sigma(\omega)=\frac{{\rm i}\sigma_0\gamma}{\omega}.
\label{eq11}
\end{equation}
\noindent
Thus, the plasma model is characterized by pure imaginary
conductivity. In the opposite limit $\omega\ll\gamma$
the unity in both numerator and denominator of (\ref{eq10})
dominate over $\omega/\gamma$ and $\omega^2/\gamma^2$
leading to $\sigma(\omega)=\sigma_0$. This converts the
dielectric permittivity of the Drude model (\ref{eq4})
in the dielectric permittivity of the normal skin effect
(\ref{eq9}) characterized by real conductivity of conduction
electrons $\sigma_0$.

The total current in the framework of the Drude model (\ref{eq4})
is given by
\begin{eqnarray}
&&
\mbox{\boldmath$j$}_{\rm tot}(\mbox{\boldmath$r$},t)=
\mbox{Re}\left[-\frac{{\rm i}\omega}{4\pi}\varepsilon_D(\omega)
\mbox{\boldmath$E$}_{0}(\mbox{\boldmath$r$})
e^{-{\rm i}\omega t}\right]
\label{eq12} \\
&&
\phantom{aaa}=\frac{\omega}{4\pi}\left(1-
\frac{\omega_p^2}{\omega^2+\gamma^2}\right)
\mbox{Im}\left[
\mbox{\boldmath$E$}_{0}(\mbox{\boldmath$r$})
e^{-{\rm i}\omega t}\right]
+\frac{\sigma_0\gamma^2}{\omega^2+\gamma^2}
\mbox{Re}\left[
\mbox{\boldmath$E$}_{0}(\mbox{\boldmath$r$})
e^{-{\rm i}\omega t}\right].
\nonumber
\end{eqnarray}
\noindent
The first term on the right-hand side of this equation has
the meaning of the displacement current, whereas the second
term, in accordance to (\ref{eq6}), is proportional to the
physical electric field 
$\mbox{\boldmath$E$}=\mbox{\boldmath$E$}(\mbox{\boldmath$r$},t)$ and
describes the real current of conduction electrons. Under the
condition $\gamma\ll\omega<\omega_p$, i.e., in the region
of infrared optics, the first term dominates. This is the
displacement current of the plasma model with a pure imaginary
conductivity (\ref{eq12}). Under the opposite condition
$\omega\ll\gamma$, i.e., in the region of the normal skin
effect,
the second term on the right-hand side of (\ref{eq12}),
i.e., the real physical current of conduction electrons
dominates.

After the above discussion on the derivation of the Drude
model we now return to the role played by the finite size
of the plates. Let us consider plane waves of zero-point
oscillations and thermal photons in between the plates
incident on their interior boundary surfaces. It is common
knowledge (see, e.g., \cite{33,34}) that charge carriers in 
a conductor are moving in response to electromagnetic
oscillations. For plates of finite area the application
condition for the derivation of the Drude model is formally
violated because the extension of the oscillation wave fronts is
much larger than the size of any conceivable plates.
However, if the frequencies of oscillations are high enough
(recall that at room temperature the first Matsubara
frequency is equal to $\xi_1\approx 2.47\times 10^{14}\,$rad/s,
and all others with $l\geq 1$ are, respectively, higher)
there is no accumulation of charges on the side boundary 
surfaces of 
finite metal plates. First, at so high frequencies the real
current of conduction electrons [given by the second term on
the right-hand side of (\ref{eq12})] is small in comparison
with the displacement current. Second, a high-frequency
electric field quickly changes its direction. As a result,
electric charges which are accumulated on the sides of a plate
change their sign many times during any 
reasonable time of force
measurement. This leads to practically zero mean surface
charge. Thus, equations   (\ref{eq4}), (\ref{eq7}), (\ref{eq8})
and (\ref{eq12}) remain macroscopically valid.

The situation changes drastically when the contribution from
the zero Matsubara frequency $\xi_0=0$ is considered. The plane
wave of zero frequency should be understood as limit of plane
waves with some low frequencies $\xi$ in the case that
$\xi\to 0$. As was mentioned above, the extension of a wave
front far exceeds the size of the plates. If it is remembered that
the period of the wave of vanishing frequency goes to infinity,
the Casimir plates are found in practically constant electric
and magnetic fields. As is described in textbooks on classical
electrodynamics (see, e.g., \cite{33,34}), in a quasistatic case
the propagation
direction of a plane wave inside a metal is approximately
perpendicular to its surface independently of the angle of
incidence. Thus, a short-lived current which arises under the
influence of a constant electric field in the plane of plates
immediately gives rise to some nonzero surface charge densities
$\rho$ of opposite signs accumulated on opposite sides of the
plates. The electric field generated by these charges precisely
compensates the electric field of zero frequency inside a metal.
As a result, the electric field inside the metal is exactly
equal to zero \cite{33,34}. As to the space between the plates,
the resulting field there is the superposition of an approximately 
constant field of external sources and of the field generated by
the charge distribution on the plate sides.

From what has been said, it appears that Maxwell equations
(\ref{eq5}) and all consequences obtained from them are not
applicable in the case of plane waves of zero frequency. In that
case for finite plates not only a nonzero induced current must
be taken into account but also a nonzero induced charge density 
generated by this current which,
however, is omitted in (\ref{eq5}). As was
noted by Parsegian (see \cite{28}, p.254), 
``conductors must be considered
case by case corresponding to the limitations imposed by
boundary surfaces.'' Here we demonstrate that these limitations 
arise from the substitution of the Drude dielectric function
(which is obtained for unbounded medium) in the zero-frequency
term of the Lifshitz formula. Such substitution is in contradiction 
with electrodynamics, because, as was shown above, the Drude model
admits a nonzero current of conduction electrons, whereas
electrodynamics assertains that the current of conduction electrons
inside a finite metal plate placed in a plane wave of zero
frequency must be equal to zero.

We emphasize that the time interval during which charges on the
sides of finite plates are accumulated and the total electric
field in a metal turns into zero is extremely short. To make sure
that this is the case we describe the dynamic process of charge 
accumulation on the sides of the plates by a simple model:
\begin{equation}
\frac{d\rho(t)}{dt}=\sigma_0E_{\rm tot}(t),
\qquad
E_{\rm tot}(t)=E-E_{\rho}(t),
\label{eq13}
\end{equation}
\noindent
where $E$ is the constant electric field along the plates
and $E_{\rho}(t)$ is the field produced by the surface charge
density till the moment $t$. For the order of magnitude
estimation it is sufficient to represent $E_{\rho}(t)$ as the
field in a plane capacitor: $E_{\rho}(t)=4\pi\rho(t)$.
Then we arrive at the solution
\begin{equation}
\rho(t)=\frac{E}{4\pi}\left(1-e^{-4\pi\sigma_0t}\right).
\label{eq14}
\end{equation}
\noindent
As an example, for Au it holds 
$4\pi\sigma_0=3.5\times10^{18}\,\mbox{s}^{-1}$ and, thus, even 
after a very short time lapse of $t=10^{-18}\,$s, $\rho(t)$
practically achieves the maximum value
$\rho(\infty)=E/(4\pi)$. Then from equation (\ref{eq13}) it
follows that the total field inside a metal vanishes,
$E_{\rm tot}(\infty)=0$, as it should be.

Thus, the substitution of the Drude dielectric function in
the zero-frequency term of the Lifshitz formula is self-contradictory.
As was recalled in Introduction, for the Drude model
it holds $r_{\rm TE}(0,k_{\bot})=0$.
This means that the TE field of zero frequency completely
penetrates into a metal. For metal plates of finite size this
inavoidably leads to instant accumulation of induced charges
on the plate sides and vanishing of both an electric field and
a current inside a metal. However, the Lifshitz formula is derived for
neutral plates without any nonzero surface charge densities. At the
same time, the Drude model admits the presence of a nonzero induced 
current. Because of this, it is not surprising that the Lifshitz
theory in combination with the Drude model violates the Nernst heat
theorem for perfect crystal lattices \cite{7,8} and was found to
be in contradiction with several experiments \cite{14,15,16,17}.
If metal plates were really infinite (as is formally suggested
in the derivation of the Lifshitz formula) the Drude model would
be applicable including the zero-frequency term. This, however, is an
unphysical case and it cannot be considered as a closed system where 
the laws of thermodynamics must be valid.

The above discussion uses the formulation of the Lifshitz formula
(\ref{eq1}) in terms of the imaginary frequency axis. However,
direct computations using the formulation in terms of real
frequencies show \cite{35} that the region of sufficiently low
real frequencies results in precisely the same contribution
to the Casimir free energy as does the zero Matsubara frequency.
As a result,
all above conclusions are equally applicable to the contribution
into the free energy from the zero-frequency term in the
Matsubara formulation and to the equivalent contribution from
low real frequencies in the formalism of real frequency axis.

By contrast with the Drude model, the plasma dielectric function
(\ref{eq11}) does not lead to a real current of conduction electrons 
and does not result in accumulation of charges on the side surfaces 
of finite metal plates. The free electron plasma model in combination 
with the Lifshitz formula satisfies the Nernst heat theorem
\cite{7,8}. However, as was mentioned in the Introduction, it is in 
disagreement with short separation experiments on the measurement
of the Casimir force. Below we demonstrate that the generalized
plasma-like permittivity \cite{27}, which is in agreement with all
experiments performed up to date also satisfies the requirements
of thermodynamics. Thus, it is becoming the best known candidate
for the adequate description of metals in the framework of
the Lifshitz theory.

\section{Thermodynamic test for the generalized plasma-like dielectric
permittivity}

The generalized plasma-like dielectric permittivity can be presented in
the form \cite{27}
\begin{equation}
\varepsilon(\omega)=1-\frac{\omega_p^2}{\omega^2}+A(\omega),
\label{eq15}
\end{equation}
\noindent
where the additional term $A(\omega)$ takes into account the interband
transitions of core electrons. Explicitly it is given by
\begin{equation}
A(\omega)=\sum\limits_{j=1}^{K}
\frac{f_j}{\omega_j^2-\omega^2-{\rm i}g_j\omega},
\label{eq16}
\end{equation}
\noindent 
where $\omega_j\neq 0$ are the resonant frequencies of oscillators
describing the core electrons, $g_j$ are the respective relaxation
parameters, $f_j$ are the oscillator strengths and $K$ is the
number of oscillators. The values of oscillator parameters for
different materials can be found in \cite{28}. Recently the precise
determination of these parameters for Au was performed in \cite{17}.
Note that the generalized plasma model does not include relaxation
of the free conduction electrons. The latter are described by an
oscillator with zero resonant frequency, $\omega_0=0$, which is
not contained in (\ref{eq16}) but is explicitly included in (\ref{eq15})
with $g_0=0$ and $f_0=\omega_p^2$.
Thus, similar to the usual plasma model (\ref{eq11}), the generalized
plasma-like permittivity (\ref{eq15}), (\ref{eq16}) admits only the 
displacement current and does not allow for the accumulation of
charges on the sides of finite plates. Because of this, the generalized
plasma-like permittivity is compatible with the Lifshitz formula
which is derived for neutral plates with zero charge distributions.

In \cite{27} it was shown that the permittivity (\ref{eq15}), (\ref{eq16})
precisely satisfies the Kramers-Kronig relations. Here we prove 
that the Lifshitz formula combined with the generalized plasma-like
permittivity is in agreement with the Nernst heat theorem  and thus
withstands the thermodynamic test.

To find the asymptotic behavior of the Casimir free energy and entropy
at low temperature, we first present equations (\ref{eq1}), (\ref{eq2})
and (\ref{eq15}), (\ref{eq16}) in terms of the following dimensionless
parameters
\begin{eqnarray}
&&
\tilde{\omega}_p=\frac{\omega_p}{\omega_c}\equiv\frac{1}{\alpha},
\qquad
\zeta_l=\frac{\xi_l}{\omega_c}\equiv\tau l,
\qquad
y=\sqrt{4a^2k_{\bot}^2+\zeta_l^2},
\nonumber \\
&&
\gamma_j=\frac{\omega_c^2}{\omega_j^2},
\qquad
\delta_j=\frac{\omega_c g_j}{\omega_j^2},
\qquad
C_j=\frac{f_j}{\omega_j^2},
\label{eq17}
\end{eqnarray}
\noindent
where $\omega_c\equiv c/(2a)$ is the so-called characteristic
frequency of the Casimir effect. In terms of new variables, the
Lifshitz formula (\ref{eq1}) takes the form
\begin{eqnarray}
&&
{\cal F}(a,T)=\frac{\hbar c\tau}{32\pi^2a^3}\sum\limits_{l=0}^{\infty}
\left(1-\frac{1}{2}\delta_{0l}\right)
\int_{\zeta_l}^{\infty}y\,dy
\nonumber \\
&&\phantom{aaa}
\times\left\{\ln\left[1-r_{\rm TM}^2(\zeta_l,y)
e^{-y}\right]+
\ln\left[1-r_{\rm TE}^2(\zeta_l,y)
e^{-y}\right]\right\}.
\label{eq18}
\end{eqnarray}
\noindent 
The reflection coefficients (\ref{eq2}) are given by
\begin{eqnarray}
&&
r_{\rm TM}(\zeta_l,y)=
\frac{(\varepsilon_l^2-1)(y^2-\zeta_l^2)}{(\varepsilon_l+1)y^2+
(\varepsilon_l-1)\zeta_l^2+2\varepsilon_lyh_l(y)
\coth\left[\frac{d}{2a}h_l(y)\right]},
\nonumber \\
&&
r_{\rm TE}(\zeta_l,y)=
\frac{(\varepsilon_l-1)\zeta_l^2}{2y^2+
(\varepsilon_l-1)\zeta_l^2+2yh_l(y)
\coth\left[\frac{d}{2a}h_l(y)\right]},
\label{eq19}
\end{eqnarray}
\noindent
where
\begin{equation}
h_l(y)=\left[y^2+(\varepsilon_l-1)\zeta_l^2\right]^{1/2}.
\label{eq20}
\end{equation}
\noindent
The generalized plasma-like dielectric permittivity along
the imaginary frequency axis can be presented as
\begin{equation}
\varepsilon_l=\varepsilon(i\zeta_l)=1+
\frac{\tilde{\omega}_p^2}{\zeta_l^2}+A_l=
1+\frac{1}{\alpha^2\zeta_l^2}+A_l,
\label{eq21}
\end{equation}
\noindent
where
\begin{equation}
A_l=A(\zeta_l)=\sum\limits_{j=1}^{K}
\frac{C_j}{1+\gamma_j\zeta_l^2+\delta_j\zeta_l}.
\label{eq22}
\end{equation}

Using the Abel-Plana formula \cite{3,5}
\begin{equation}
\sum\limits_{l=0}^{\infty}
\left(1-\frac{1}{2}\delta_{0l}\right)F(l)=
\int_{0}^{\infty}F(t)dt+i\int_{0}^{\infty}dt
\frac{F({\rm i}t)-F(-{\rm i}t)}{e^{2\pi t}-1},
\label{eq23}
\end{equation}
\noindent
where $F(z)$ is an analytic function in the right half of
the complex plane, we can rearrange (\ref{eq18}) to the form
\begin{equation}
{\cal F}(a,T)=E(a)+\Delta{\cal F}(a,t).
\label{eq24}
\end{equation}
\noindent
Here, the energy of the Casimir interaction at zero temperature
is given by
\begin{equation}
E(a)=\frac{\hbar c}{32\pi^2a^3}\int_{0}^{\infty}d\zeta
\int_{\zeta}^{\infty}f(\zeta,y)dy
\label{eq25}
\end{equation}
\noindent
and the function $f(\zeta,y)$ is defined as
\begin{equation}
f(\zeta,y)=y\ln\left[1-r_{\rm TM}^2(\zeta,y)e^{-y}\right]+
y\ln\left[1-r_{\rm TE}^2(\zeta,y)e^{-y}\right].
\label{eq26}
\end{equation}
\noindent
The thermal correction to the Casimir energy is expressed as
follows:
\begin{equation}
\Delta{\cal F}(a,T)=\frac{i\hbar c\tau}{32\pi^2a^3}
\int_{0}^{\infty}dt
\frac{F({\rm i}\tau t)-F(-{\rm i}\tau t)}{e^{2\pi t}-1},
\label{eq27}
\end{equation}
\noindent
where
\begin{equation}
F(x)=\int_{x}^{\infty}dy\,f(x,y).
\label{eq28}
\end{equation}
\noindent
The behavior of the thermal correction (\ref{eq27}), (\ref{eq28})
at low temperature will be the subject of our further
consideration.

Perturbation expansion can be performed in analogy to papers
\cite{7,8,36,37}. At first, we expand the reflection coefficients
(\ref{eq19}) with $\zeta_l$ replaced by $\zeta$
in powers of parameter $\alpha$ defined in (\ref{eq17})
preserving all powers up to the fourth inclusive. The parameter
$\alpha$ can be identically presented as $\alpha=\lambda_p/(4\pi a)$,
where $\lambda_p$ is the plasma wavelength. This means that
$\alpha\ll 1$ at all separation distances between the plates
larger than $\lambda_p$. As is seen from (\ref{eq26}), it is more
convenient to expand the logarithmic functions containing the
reflection coefficients in (\ref{eq26}) multiplied by the
variable $y$. The results are:
\begin{eqnarray}
&&
y\ln\left[1-r_{\rm TM}^2(\zeta,y)e^{-y}\right]=
y\ln\left(1-e^{-y}\right)+\alpha\frac{4\zeta^2}{e^{y}-1}
-\alpha^2\frac{8e^y\zeta^4}{y\left(e^y-1\right)^2}
\nonumber \\
&&\phantom{a}
+\alpha^3\frac{2\zeta^2
\left\{2\zeta^4\left(3e^y+1\right)^2+3\left(e^y-1\right)^2y^2
\left[y^2-2\zeta^2-\zeta^2A(\zeta)
\right]\right\}}{3y^2\left(e^{y}-1\right)^3}
\nonumber \\
&&\phantom{a}
-\alpha^4\frac{8e^y\zeta^4\left\{2\zeta^4\left(e^y+1\right)^2+
\left(e^y-1\right)^2y^2
\left[y^2-2\zeta^2-\zeta^2A(\zeta)
\right]\right\}}{y^3\left(e^y-1\right)^4}\, ,
\nonumber \\
&&
\label{eq29} \\
&&
y\ln\left[1-r_{\rm TE}^2(\zeta,y)e^{-y}\right]=
y\ln\left(1-e^{-y}\right)+\alpha\frac{4y^2}{e^{y}-1}
-\alpha^2\frac{8y^3e^y}{\left(e^y-1\right)^2}
\nonumber \\
&&\phantom{a}
+\alpha^3\frac{2y^2\left[-3\left(e^y-1\right)^2\zeta^2A(\zeta)+
y^2\left(15e^{2y}+18e^{y}-1\right)\right]}{3\left(e^y-1\right)^3}
\nonumber \\
&&\phantom{a}
-\alpha^4\frac{8y^3e^y\left[-\left(e^y-1\right)^2\zeta^2A(\zeta)+
y^2\left(e^{2y}+6e^y+1\right)\right]}{\left(e^y-1\right)^4}\, .
\nonumber
\end{eqnarray}

It is significant that these expansions do not 
depend on $d$ (the thickness of the plates) contained in (\ref{eq19}).
This is because the factor in the denominator of (\ref{eq19}),
\begin{eqnarray}
&&
\coth\left[\frac{d}{2a}h_l(y)\right]=
\coth\left(\frac{d}{2a}\sqrt{y^2+\frac{1}{\alpha^2}+
A_l\zeta_l^2}\right)
\label{eq30} \\
&&\phantom{aa}
=\frac{1+\exp\left(-\frac{d}{a\alpha}\sqrt{1+\alpha^2y^2+
\alpha^2A_l\zeta_l^2}\right)}{1-
\exp\left(-\frac{d}{a\alpha}\sqrt{1+\alpha^2y^2+
\alpha^2A_l\zeta_l^2}\right)},
\nonumber
\end{eqnarray}
\noindent
behaves asymptotically as
\begin{equation}
1+2\exp\left(-\frac{d}{a\alpha}\right)+\,\ldots
\label{eq31}
\end{equation}
\noindent
when $\alpha$ goes to zero. Thus, this factor could only contribute
exponentially small terms in the expansion (\ref{eq29})
providing the plate thickness $d$ is much larger than the
penetration depth of electromagnetic oscillations into the metal
[recall that $2a\alpha=\lambda_p/(2\pi)$]. Under this condition
the perturbation expansions (\ref{eq29}) are common for two
semispaces and for two plates of finite thickness. We note also
that terms in (\ref{eq29}) of order $\alpha^0$, $\alpha$ and
$\alpha^2$ do not contain contributions from the core electrons.
They are the same as for the usual free electron plasma model
(\ref{eq11}). The contributions from the core electrons are
contained only in the terms of order $\alpha^3$ and $\alpha^4$
in (\ref{eq29}).

The parameter $\tau$ defined in (\ref{eq17}) can be identically
represented as
\begin{equation}
\tau=2\pi\frac{T}{T_{\rm eff}}, \qquad
k_BT_{\rm eff}\equiv\frac{\hbar c}{2a}.
\label{eq32}
\end{equation}
\noindent
Here $T_{\rm eff}$ is the so-called effective temperature. For
example, at a separation of $a=1\,\mu$m 
$T_{\rm eff}\approx 1145\,$K. Below we will consider the limiting
case of low temperatures $T\ll T_{\rm eff}$.
The contribution from the terms of order $\alpha^0$, $\alpha$ and
$\alpha^2$ in (\ref{eq29}) into the thermal correction (\ref{eq26})
was found in \cite{36,37} where the usual free electron plasma
model was considered. This contribution is given by
\begin{eqnarray}
&&
\Delta{\cal F}_p(a,T)=-\frac{\hbar c}{32\pi^2a^3}\left\{
\vphantom{\left[\frac{\zeta(3)}{\pi^2}\tau^3\right]}
\frac{\zeta(3)}{4\pi^2}\tau^3-\frac{1}{360}\tau^4\right.
\label{eq33}\\
&&\phantom{aa}\left.
+\alpha\left[\frac{\zeta(3)}{\pi^2}\tau^3
-\frac{1}{45}\tau^4 \right]-
\alpha^2\frac{6\zeta(5)}{\pi^2}\tau^5\right\},
\nonumber
\end{eqnarray}
where $\zeta(z)$ is the Riemann zeta function. As was shown
in \cite{37}, the terms in (\ref{eq33}) of order $\alpha^0$ and
$\alpha$ do not contain corrections of order $\tau^n$ with
$n\geq 5$. They contain only the exponentially small
corrections of order exp$(-2\pi/\tau)$.

Now we deal with the terms of order $\alpha^3$ and $\alpha^4$
in (\ref{eq29}) which contain the contributions from the core
electrons. From (\ref{eq26}), (\ref{eq28}) and (\ref{eq29}) the
respective functions $F^{(3)}(x)$ and $F^{(4)}(x)$ are
given by
\begin{eqnarray}
&&
F^{(3)}(x)=-2\alpha^3\left\{
\vphantom{\int_{x}^{\infty}
\frac{\left(3e^{y}+1\right)^2}{y^3\left(e^y-1\right)^3}}
\biggl[A(x)-1\biggr]x^2\int_{x}^{\infty}
\frac{y^2dy}{e^y-1}
\right.
\nonumber \\
&&\phantom{aaaa}
-\frac{1}{3}\int_{x}^{\infty}
\frac{y^4\left(15e^{2y}+18e^{y}-1\right)}{\left(e^y-1\right)^3}
+\biggl[A(x)+2\biggr]x^4\int_{x}^{\infty}
\frac{dy}{e^y-1}
\nonumber \\
&&\phantom{aa}\left.
-\frac{2}{3}x^6\int_{x}^{\infty}
\frac{\left(3e^{y}+1\right)^2}{y^3\left(e^y-1\right)^3}dy
\right\},
\nonumber \\
&& \label{eq34} \\
&&
F^{(4)}(x)=8\alpha^4\left\{
\vphantom{\int_{x}^{\infty}
\frac{\left(3e^{y}+1\right)^2}{y^3\left(e^y-1\right)^3}}
A(x)x^2\int_{x}^{\infty}
\frac{y^3e^{y}dy}{\left(e^y-1\right)^2}
\right.
\nonumber \\
&&\phantom{aa}
-\int_{x}^{\infty}
\frac{y^5\left(e^{2y}+6e^{y}+1\right)e^{y}dy}{\left(e^y-1\right)^4}
+\biggl[A(x)+2\biggr]x^6\int_{x}^{\infty}
\frac{e^{y}dy}{y\left(e^y-1\right)^2}
\nonumber \\
&&\phantom{aa}\left.
-x^4\int_{x}^{\infty}
\frac{ye^{y}dy}{y^3\left(e^y-1\right)^2}
-2x^8\int_{x}^{\infty}
\frac{e^{y}\left(e^{y}+1\right)^2}{y^3\left(e^y-1\right)^4}dy
\right\}.
\nonumber 
\end{eqnarray}

Calculating all integrals in (\ref{eq34}) as asymptotic
expansions at small $x$ (see Appendix for details) we arrive at
\begin{eqnarray}
&&
F^{(3)}(i\tau t)-F^{(3)}(-i\tau t)=-2{\rm i}\alpha^3\left[
\vphantom{\left(\sum\limits_{j=1}^{K}C_j\right)}
2\tau^3t^3\zeta(3)\sum\limits_{j=1}^{K}C_j\delta_j
\right.
\nonumber \\
&&\phantom{aaaaaaaaaaaaaaaaaa}
+\left.
\pi\tau^4t^4\left(\sum\limits_{j=1}^{K}C_j+2\right)\right],
\label{eq35} \\
&&
F^{(4)}({\rm i}\tau t)-F^{(4)}(-{\rm i}\tau t)=8{\rm i}\alpha^4\left[
8\tau^3t^3\zeta(3)\sum\limits_{j=1}^{K}C_j\delta_j+
\pi\tau^4t^4\right].
\nonumber
\end{eqnarray}
\noindent
The terms omitted in (\ref{eq35}) are of order $\tau^5$.

Substituting (\ref{eq35}) in (\ref{eq27}) and integrating with
respect to $t$, we obtain the contribution to the thermal
correction from the terms of order $\alpha^3$ and $\alpha^4$:
\begin{eqnarray}
&&
\Delta{\cal F}_g(a,T)=-\frac{\hbar c}{32\pi^2a^3}\left\{
-\alpha^3\left[\frac{\zeta(3)}{60}\sum\limits_{j=1}^{K}C_j\delta_j 
\,\tau^4+
\frac{3\zeta(5)}{2\pi^4}\left(\sum\limits_{j=1}^{K}C_j+2\right)
\,\tau^5\right]\right.
\nonumber \\
&&\phantom{aaa}\left.
+\alpha^4\left[\frac{4\zeta(3)}{15}\sum\limits_{j=1}^{K}C_j\delta_j 
\,\tau^4+
\frac{6\zeta(5)}{\pi^4}\tau^5\right]\right\}.
\label{eq36}
\end{eqnarray}

The total Casimir free energy computed using the generalized
plasma-like permittivity can be now found from (\ref{eq24}),
(\ref{eq33}) and (\ref{eq36}):
\begin{equation}
{\cal F}(a,T)=E(a)+\Delta{\cal F}_p(a,T)+\Delta{\cal F}_g(a,T).
\label{eq37}
\end{equation}
\noindent
Taking into account (\ref{eq32}), it can be represented in
the form
\begin{eqnarray}
&&
{\cal F}(a,T)=E(a)-\frac{\hbar c\zeta(3)}{16\pi a^3}
\left(\frac{T}{T_{\rm eff}}\right)^3\left\{
\vphantom{\left(\frac{T}{T_{\rm eff}}\right)^2
\left[\left(\sum\limits_{j=1}^{K}C_j+2\right)
-\frac{\alpha^2}{\pi^2}\right]}
1+4\alpha \right.
\nonumber \\
&&\phantom{aaa}
-\frac{\pi^3}{45\zeta(3)}\frac{T}{T_{\rm eff}}
\left(1+8\alpha+6\zeta(3)\alpha^3\sum\limits_{j=1}^{K}C_j\delta_j
-96\zeta(3)\alpha^4\sum\limits_{j=1}^{K}C_j\delta_j\right)
\nonumber \\
&&\phantom{aaa}
\left.
-\frac{96\pi^2\zeta(5)}{\zeta(3)}
\left(\frac{T}{T_{\rm eff}}\right)^2
\alpha^2\left[1+\frac{\alpha}{4\pi^2}
\left(\sum\limits_{j=1}^{K}C_j+2\right)
-\frac{\alpha^2}{\pi^2}\right]\right\}.
\label{eq38}
\end{eqnarray}
\noindent
Here one can see that the free energy calculated using the 
generalized plasma-like permittivity contains the correction
of order $(T/T_{\rm eff})^4$ not only in the terms of order
$\alpha^0$ and $\alpha$ (as in the usual plasma model)
but also in the third and fourth order expansion 
terms in $\alpha$. In the usual plasma model 
the terms of order $\alpha^3$ and $\alpha^4$ contain the
thermal corrections only of order of $(T/T_{\rm eff})^5$ and
higher \cite{36}. To estimate the relative role of the additional
terms arising due to the use of the generalized plasma-like
permittivity, one can use the parameters of oscillator terms
in (\ref{eq22}) for Au \cite{17}. This results in
\begin{equation}
\nonumber
\sum\limits_{j=1}^{6}C_j=6.3175, \qquad
\sum\limits_{j=1}^{6}C_j\delta_j=\left\{
\begin{array}{ll}
0.272, & a=200\,\mbox{nm}, \\
0.109, & a=500\,\mbox{nm}.
\end{array}
\right.
\nonumber
\end{equation}

From (\ref{eq38}) it is easy to find the asymptotic behavior of
the Casimir entropy
\begin{equation}
S(a,T)=-\frac{\partial{\cal F}(a,T)}{\partial T}
\label{eq39}
\end{equation}
\noindent
at low temperature. The result is
\begin{eqnarray}
&&
S(a,T)=\frac{3\zeta(3)k_B}{8\pi a^2}
\left(\frac{T}{T_{\rm eff}}\right)^2\left\{
\vphantom{\left(\frac{T}{T_{\rm eff}}\right)^2
\left[\left(\sum\limits_{j=1}^{K}C_j+2\right)
-\frac{\alpha^2}{\pi^2}\right]}
1+4\alpha \right.
\nonumber \\
&&\phantom{aaa}
-\frac{4\pi^3}{135\zeta(3)}\frac{T}{T_{\rm eff}}
\left(1+8\alpha+6\zeta(3)\alpha^3\sum\limits_{j=1}^{K}C_j\delta_j
-96\zeta(3)\alpha^4\sum\limits_{j=1}^{K}C_j\delta_j\right)
\nonumber \\
&&\phantom{aaa}
\left.
-\frac{160\pi^2\zeta(5)}{\zeta(3)}
\left(\frac{T}{T_{\rm eff}}\right)^2
\alpha^2\left[1+\frac{\alpha}{4\pi^2}
\left(\sum\limits_{j=1}^{K}C_j+2\right)
-\frac{\alpha^2}{\pi^2}\right]\right\}.
\label{eq40}
\end{eqnarray}

As is seen from (\ref{eq40}),
\begin{equation}
S(a,T)\to 0 \quad \mbox{when} \quad T\to 0,
\label{eq41}
\end{equation}
\noindent
i.e., the entropy goes to zero (and remains positive)
when the temperature vanishes. This means that the Nernst
heat theorem is satisfied and the Lifshitz theory combined
with the generalized plasma-like dielectric permittivity
withstands the thermodynamic test.

\section{Conclusions and discussion}

In the foregoing we have continued the elaboration of new
theoretical approach to the thermal Casimir force based on
the Lifshitz formula combined with the generalized
plasma-like dielectric permittivity \cite{27}.
In the first part of the paper (Section 2) the physical
reasons were presented why the Drude dielectric function
is not applicable in the case of finite metallic plates.
It was shown that for the validity of the Drude
model the nonzero current of conduction electrons must
exist, whereas the surface charge densities must be equal to zero.
Both these conditions are shown to be violated when the plane
wave of electromagnetic oscillations of vanishing frequency
falls on a finite metal plate. In this case the electric field 
and current of conduction electrons practically instantaneously
turn into zero. This is accompanied by the accumulation of
charges on the sides of plates. Not only the Drude dielectric
function, but also the Lifshitz formula are not applicable
to this physical situation. On the contrary, the generalized
plasma-like permittivity leads to only a displacement current
and does not result in the accumulation of surface charges.
The obtained results furnish insights into the long-debated
problem why the Lifshitz theory combined with the Drude
dielectric function results in contradictions with
thermodynamics and experiment. It becomes clear also why the 
generalized plasma-like permittivity, which does not include the
relaxation processes of conduction electrons, is consistent
with all available measurements of the Casimir force at both
short and large separations.

Recent paper \cite{41a} also argues that the finite size  effects
of the conductors may play an important role in the problem of
the thermal Casimir effect. This was illustrated in the simplified
case of two wires of finite length described by the Drude model
and interacting through the inductive coupling between Johnson
currents. If the capacitive effects associated with the end points 
of the wires are not taken into account, the thermal interaction 
between the wires leads to the violation of the Nernst heat theorem.
If the capacitive effects were taken into consideration, the
agreement with thermodynamics is restored \cite{41a}.

To conclusively establish the applicability of the generalized
plasma-like permittivity in the theory of the thermal Casimir
force between metals, in Section 3 we have performed
the thermodynamic test of this model. We have analytically found 
the asymptotic behavior   of both  the Casimir free energy and
Casimir entropy at low temperature. This was done using the
perturbation theory in two small parameters. The obtained new
expressions generalize the previously known ones (found for the
usual free electron plasma model which does not take dissipation
into account). When the oscillator parameters describing the core
electrons go to zero, the newly obtained expressions for the
Casimir free energy and entropy go into the ones found for the
usual plasma model. The Casimir entropy at low temperature
derived using the generalized plasma-like permittivity is positive
and takes zero value at zero temperature. Thus, the Nernst heat
theorem is satisfied.

To conclude, the generalized plasma-like permettivity provides
a good basis in agreement with thermodynamics and experiment
for the description of the thermal Casimir force between  
metallic plates of finite size using the standard Lifshitz theory.
A more fundamental approach to the resolution of this problem 
would require, in accordance with Parsegian's insight \cite{28},
the consideration from the very beginning 
of finite plates and charging of their boundary
surfaces. This, however, goes far beyond the scope of the
Lifshitz theory.

\section*{Acknowledgments}
 G.L.K. and V.M.M. are
grateful to the Center of Theoretical Studies and Institute
for Theoretical Physics, Leipzig University for kind
hospitality.
This work  was supported by Deutsche Forschungsgemeinschaft,
Grant No.~436\,RUS\,113/789/0--3.
\appendix
\section*{Appendix}
\setcounter{section}{1}
Here we derive equations (\ref{eq35}), where functions $F^{(3)}(x)$
and $F^{(4)}(x)$ are defined in (\ref{eq34}).

The first integral in the definition of $F^{(3)}(x)$ converges when
$x\to 0$ and can be calculated as
\begin{eqnarray}
&&
I_1^{(3)}(x)\equiv\int_{x}^{\infty}
\frac{y^2dy}{e^{y}-1}= 2\mbox{Li}_3\left(e^{-x}\right)+
2x\mbox{Li}_2\left(e^{-x}\right) -x^2\ln\left(1-e^{-x}\right)
\nonumber \\
&&\phantom{aaa}
=2\zeta(3)-\frac{x^2}{2}+\frac{x^3}{6}-\frac{x^4}{48}+
\mbox{O}(x^6),
\label{a1}
\end{eqnarray}
\noindent
where Li${}_n(z)$ is the polylogarithm function \cite{41b}.
Expanding the function $A(x)$ defined in (\ref{eq22}) in powers of $x$
and using (\ref{a1}) we arrive at
\begin{eqnarray}
&&
\left[A(x)-1\right]x^2I_1^{(3)}(x)
=-2\zeta(3)x^2+\frac{x^4}{2}+2\zeta(3)\sum\limits_{j=1}^{K}C_jx^2
-2\zeta(3)\sum\limits_{j=1}^{K}C_j\delta_jx^3
\nonumber \\
&&\phantom{aaa}
-\left[\frac{1}{2}\sum\limits_{j=1}^{K}C_j-
2\zeta(3)\sum\limits_{j=1}^{K}C_j\delta_j^2+
2\zeta(3)\sum\limits_{j=1}^{K}C_j\gamma_j\right]x^4
+\mbox{O}(x^5),
\label{a2}
\end{eqnarray}
\noindent
From (\ref{a2}), only the term proportional to $x^3$ contributes
to (\ref{eq35}).

The second integral in the definition of $F^{(3)}(x)$ in (\ref{eq34})
also converges when $x\to 0$. It can be found in the form
\begin{eqnarray}
&&
I_2^{(3)}(x)\equiv\int_{x}^{\infty}
\frac{y^4\left(15e^{2y}+18e^{y}-1\right)dy}{\left(e^{y}-1\right)^3}
=15\sum\limits_{n=1}^{\infty}n^2\int_{x}^{\infty}y^4e^{-ny}dy
\nonumber \\
&&\phantom{aaa}
+\int_{x}^{\infty}
\frac{y^4e^{-2y}dy}{\left(1-e^{-y}\right)^2}
=-16\frac{x^2}{2}+\frac{x^4}{5}+
\mbox{O}(x^5).
\label{a3}
\end{eqnarray}
\noindent
This does not contribute to (\ref{eq35}) in the perturbation orders
under consideration.

The third integral in $F^{(3)}(x)$,
\begin{equation}
I_3^{(3)}(x)\equiv\int_{x}^{\infty}
\frac{e^{-y}dy}{1-e^{-y}}=-\ln\left(1-e^{-x}\right),
\label{a4}
\end{equation}
\noindent
diverges when $x$ goes to zero.
It should be, however, multiplied by $\left[A(x)+2\right]x^4$ with 
the result
\begin{equation}
\left[A(x)+2\right]x^4I_3^{(3)}(x)=
-\left(\sum\limits_{j=1}^{K}{\!}C_j+2{\!}\right){\!}x^4\ln{x}+
\sum\limits_{j=1}^{K}{\!}C_j\delta_jx^5\ln{x}+\mbox{O}(x^5).
\label{a5}
\end{equation}
\noindent
Only the first term on the right-hand side of (\ref{a5}) contributes
to (\ref{eq35}). This contribution is simply found when taken
into account that
\begin{equation}
\ln({\rm i}z)-\ln(-{\rm i}z)={\rm i}\pi.
\label{a6}
\end{equation}

The last, fourth, integral in the definition of $F^{(3)}(x)$,
also diverges when $x$ goes to zero. It can be identically
represented in the form
\begin{eqnarray}
&&
I_4^{(3)}(x)\equiv\int_{x}^{\infty}
\frac{e^{-y}\left(3+e^{-y}\right)^2}{y^2\left(1-e^{-y}\right)^3}dy
=16\int_{x}^{\infty}\frac{e^{-y}}{y^5}dy+
16\int_{x}^{\infty}\frac{e^{-y}}{y^4}dy
\nonumber \\
&&
\phantom{aaa}
+9\int_{x}^{\infty}\frac{e^{-y}}{y^3}dy+
\frac{19}{6}\int_{x}^{\infty}\frac{e^{-y}}{y^2}dy
+\frac{41}{60}\int_{x}^{\infty}\frac{e^{-y}}{y}dy
\nonumber \\
&&
\phantom{aaa}
+\int_{x}^{\infty}\frac{e^{-y}}{y^2}\left[
\frac{\left(3+e^{-y}\right)^2}{\left(1-e^{-y}\right)^3}-
\frac{16}{y^3}-\frac{16}{y^2}-\frac{9}{y}-
\frac{19}{6}-\frac{41y}{60}\right]dy.
\label{a7}
\end{eqnarray}
\noindent
The last integral on the right-hand side of (\ref{a7})
converges when $x\to 0$. It does not contribute to the 
perturbation orders of our interest after the multiplication
by $x^5$. As a result, we arrive at
\begin{eqnarray}
&&
x^6I_4^{(3)}(x)=x^6\left[
\vphantom{\frac{41}{60}}
16\Gamma(-4,x)+16\Gamma(-3,x)+9\Gamma(-2,x)
\right.
\nonumber \\
&&\phantom{aaaaaaaaa}
\left.
+\frac{19}{6}\Gamma(-1,x)+\frac{41}{60}\Gamma(-0,x)\right]
+\mbox{O}(x^6),
\label{a8}
\end{eqnarray}
\noindent
where $\Gamma(n,x)$ is the incomplete gamma function \cite{38}.
Using the identity \cite{38}
\begin{equation}
\Gamma(-n,x)=\frac{(-1)^n}{n!}\left[\Gamma(0,x)-
e^{-x}\sum\limits_{m=0}^{n-1}(-1)^m\frac{m!}{x^{m+1}}\right],
\label{a9}
\end{equation}
\noindent
where $n=1,\,2,\,\ldots\,$, and the asymptotic relation
\begin{equation}
\Gamma(0,x)=-\gamma-\ln{x}+x-\frac{x^2}{4}+\frac{x^3}{18}
+\mbox{O}(x^4)
\label{a10}
\end{equation}
\noindent
with Euler's constant $\gamma=0.577216$, we finally obtain
\begin{equation}
x^6I_4^{(3)}(x)=4x^2+\frac{x^4}{2}+\mbox{O}(x^5).
\label{a11}
\end{equation}
\noindent
This evidently does not contribute to (\ref{eq35}).

As a result, by using (\ref{a2}) and (\ref{a5}) we obtain
the first equation in (\ref{eq35}).

Now we consider the derivation of the second equation in 
(\ref{eq35}) containing the function $F^{(4)}(x)$ defined in
(\ref{eq34}). The first integral in $F^{(4)}(x)$,
\begin{equation}
I_1^{(4)}(x)\equiv\int_{x}^{\infty}
\frac{y^3e^{-y}}{\left(1-e^{-y}\right)^2}dy,
\label{a12}
\end{equation}
\noindent
converges when $x$ goes to zero. It is calculated in analogy
to (\ref{a1}). As a result the following expansion is obtained:
\begin{eqnarray}
&&
A(x)x^2I_1^{(4)}(x)=6\zeta(3)\sum\limits_{j=1}^{K}C_jx^2-
6\zeta(3)\sum\limits_{j=1}^{K}C_j\delta_jx^3
\nonumber \\
&&
\phantom{aaa}
-\left[\frac{1}{2}\sum\limits_{j=1}^{K}C_j-
6\zeta(3)\sum\limits_{j=1}^{K}C_j\delta_j^2+
6\zeta(3)\sum\limits_{j=1}^{K}C_j\gamma_j\right]x^4+
\mbox{O}(x^5).
\label{a13}
\end{eqnarray}
\noindent
The second term on the right-hand side of (\ref{a13})
contributes to (\ref{eq35}).

The second integral  in $F^{(4)}(x)$ also converges when 
$x$ goes to zero. It can be calculated similar to $I_2^{(3)}(x)$
in (\ref{a2}) and does not contain odd powers of $x$ lower than 
$x^5$. Thus, this integral does not contribute to (\ref{eq35}).

The third and fifth integrals in the definition of $F^{(4)}(x)$,
(\ref{eq34}), diverge in the limit $x\to 0$. However, by
calculating them similar to $I_4^{(3)}(x)$ in
(\ref{a7})--(\ref{a10}) and multiplying the results by $x^6$
and $x^8$, respectively, we find that
both these integrals do not contribute to (\ref{eq35}).

The fourth integral in $F^{(4)}(x)$ can be calculated as
follows:
\begin{equation}
I_4^{(4)}(x)\equiv\int_{x}^{\infty}
\frac{ye^{-y}dy}{\left(1-e^{-y}\right)^2}=
-\ln\left(1-e^{-x}\right)+\frac{xe^{-x}}{1-e^{-x}}.
\label{a14}
\end{equation}
\noindent
It diverges when $x$ goes to zero. After multiplication
by $x^4$ one obtains
\begin{equation}
x^4I_4^{(4)}(x)=x^2-x^4\ln{x}+\mbox{O}(x^6).
\label{a15}
\end{equation}
\noindent
The second term on the right-hand side of (\ref{a15}) 
contributes to (\ref{eq35}).

Using (\ref{a13}) and (\ref{a15}) we obtain
the second equation in (\ref{eq35}).


\end{document}